\documentclass[prb,letterpaper,twocolumn,showpacs,floatfix,superscriptaddress,amsmath,amsfonts,amssymb,preprintnumbers,citeautoscript]{revtex4-1}
\usepackage{times}
\usepackage{graphicx}
\usepackage{amsmath}
\usepackage{amssymb}
\usepackage[colorlinks,linkcolor=blue,citecolor=blue,urlcolor=blue,hyperindex,pdfstartview=FitH,plainpages=false]{hyperref}

\newcommand {\bisco}{Bi$_2$Sr$_2$CaCu$_2$O$_{8+\delta}$}
\newcommand {\uJcm}{$\mu$J/cm$^2$}

\begin{document}
\title{Signatures of superconductivity and pseudogap formation in non-equilibrium nodal quasiparticles revealed by ultrafast angle-resolved photoemission}

\author{Wentao Zhang}
\affiliation{Materials Sciences Division, Lawrence Berkeley National Laboratory, Berkeley, California 94720, USA}
\author{Christopher L. Smallwood}
\affiliation{Materials Sciences Division, Lawrence Berkeley National Laboratory, Berkeley, California 94720, USA}
\affiliation{Department of Physics, University of California, Berkeley, California 94720, USA}
\author{Chris Jozwiak}
\affiliation{Advanced Light Source, Lawrence Berkeley National Laboratory, Berkeley, California 94720, USA}
\author{Tristan Miller}
\affiliation{Materials Sciences Division, Lawrence Berkeley National Laboratory, Berkeley, California 94720, USA}
\affiliation{Department of Physics, University of California, Berkeley, California 94720, USA}
\author{Yoshiyuki Yoshida}
\author{Hiroshi Eisaki}
\affiliation{Electronics and Photonics Research Institute, National Institute of Advanced Industrial Science and Technology, Ibaraki 305-8568, Japan}
\author{Dung-Hai Lee}
\affiliation{Department of Physics, University of California, Berkeley, California 94720, USA}
\author{Alessandra Lanzara}
\email{alanzara@lbl.gov}
\affiliation{Materials Sciences Division, Lawrence Berkeley National Laboratory, Berkeley, California 94720, USA}
\affiliation{Department of Physics, University of California, Berkeley, California 94720, USA}
\date {\today}

\begin{abstract}

We use time- and angle-resolved photoemission to measure the nodal non-equilibrium electronic states in various dopings of \bisco.
We find that the initial pump-induced transient signal of these ungapped states is strongly affected by the onset of the superconducting gap at $T_c$, superconducting pairing fluctuations at $T_p$, and the pseudogap at $T^*$.
Moreover, $T_p$ marks a suggestive threshold in the fluence-dependent transient signal, with the appearance of a critical fluence below $T_p$ that corresponds to the energy required to break apart all Cooper pairs.
These results challenge the notion of a nodal-antinodal dichotomy in cuprate superconductors by establishing a new link between nodal quasiparticles and the cuprate phase diagram.

\end{abstract}
\maketitle

\section{Introduction}
Cuprate superconductors are known not only for extraordinarily high critical temperatures, but also for the richness of their phase diagram, where multiple energy scales associated with different electronic orders coexist at low carrier concentration and eventually merge with the critical temperature $T_c$ at higher carrier concentration.
The conventionally held wisdom is that antinodal quasiparticles shape this phase diagram. 

Indeed, in conventional superconductors the energy gap and the low-energy quasiparticle spectral weight  (i.e., the area under the quasiparticle peak) are virtually isotropic around the normal state Fermi surface.
In contrast, in high-$T_c$ cuprate superconductors superconductors the gap exhibits four nodes along the Brillouin zone diagonals (nodal direction), and the quasiparticle spectral weight is strongly momentum-dependent\cite{Shen1997}.
For example, while the quasiparticle peak in underdoped cuprates exists both above and below $T_c$ along the nodal direction, it only appears below $T_c$ along the antinodal direction\cite{Norman1998}.
This anisotropic character is assumed to derive from the ``d-wave" symmetry of the superconducting state.
Because of this sensitivity to $T_c$, the antinodal quasiparticle excitations have been regarded as carrying the information of superconductivity.
In harmony with this notion, it has been shown that the antinodal quasiparticle spectral weight scales with the critical temperature\cite{Loeser1997,Fedorov1999,Feng2000,Ding2001}, and recently it has been shown that this same spectral weight is also linked to the onset of the superconducting pair fluctuations at $T_p$\cite{Kondo2011}.
This dichotomous behavior between nodal and antinodal quasiparticles persists even above the critical temperature, in the so-called pseudogap phase, up to $T^*$.

Despite their central role in controlling most of the low energy properties of cuprate superconductors, the role of nodal quasiparticles in the superconducting transitions and more generally in shaping the cuprate phase diagram is still unclear and generally considered negligible\cite{Valla1999,Kordyuk2006,Lee2008,Kondo2009}.
This view has been challenged by a time- and angle-resolved photoemission experiment (trARPES) showing that nodal quasiparticles also respond to the superconducting transition and their spectral weight scales with the superfluid density\cite{Graf2011} and by the report of a nodeless energy gap in a very weakly doped sample\cite{Vishik2012}.

Here, we use trARPES to investigate changes in the nodal electron dynamics across a range of dopings and temperatures in the phase diagram of \bisco~(Bi2212).
We found that the initial pump-induced nodal quasiparticle population exhibits sharp features reflecting the opening of the pseudogap at $T^*$ and the onset of superconductivity at $T_c$, as well as an intermediate feature at $T_p$, between $T^*$ and $T_c$.
Below $T_p$, fluence-dependent measurements reveal a critical fluence that corresponds to the energy required to break apart all Cooper pairs, suggesting that $T_p$ is the onset temperature below which electrons begin to pair incoherently, and that it is distinct from $T^*$, which marks the onset of an independent electronic order.
To our knowledge, the present work is the first demonstration that all three of these characteristic temperatures affect the dynamics of nodal quasiparticles, and not just the dynamics of antinodal quasiparticles.

\section{Experiment}
In our trARPES experiments\cite{Smallwood2012a}, an infrared pump laser pulse (h$\nu$ = 1.48 eV) drives the sample into a nonequilibrium state, which is probed by an ultraviolet laser pulse (h$\nu$ = 5.93 eV), with repetition rate 543 kHz. The beam spot size (FWHM) of pump and probe are 100 $\mu$m and 40 $\mu$m, respectively. The delay time ($t$) between pump and probe pulses is controlled using a translation stage that varies the path length of the pump.
For $t<0$ the probe pulse arrives before the pump pulse, corresponding to an equilibrium measurement. For $t>0$ the probe pulse arrives after the pump pulse, corresponding to a non-equilibrium measurement.
The time resolution ($\sim$300 fs) and $t=0$ are determined by the cross-correlation of the pump and probe pulses as measured on polycrystalline gold with a 0.4 eV kinetic energy window centered 1.1 eV above the Fermi level.
The system is equipped with a Phoibos 150 mm hemispherical electron energy analyzer (SPECS).
The total energy resolution in the experiments is $\sim$22 meV, and the momentum resolution is $\sim$0.003 \AA$^{-1}$ at the Fermi energy.

Single-crystal samples from four different dopings of Bi2212 were measured: underdoped samples with $T_c=78$ K (UD78K) and $T_c=84$ K (UD84K); a nearly optimally doped sample with $T_c=91$ K (OP91K); and an overdoped sample with $T_c=84$ K (OD84K). We also measured an overdoped sample of Bi$_{1.7}$Pb$_{0.4}$Sr$_{1.6}$Cu$_2$O$_{6+\delta}$ (Bi2201) with $T_c=32$ K.
All single crystals were grown by the traveling solvent floating zone method. The underdoped samples were obtained by annealing the optimally doped sample in nitrogen. The overdoped Bi2212 sample was obtained by annealing the optimally doped sample in oxygen.
All the samples were cleaved \emph{in situ} in vacuum with a base pressure less than $5\times10^{-11}$ Torr.
In the temperature-dependent measurements the samples were cleaved at temperatures at or above 25 K to minimize the sample surface aging effect caused by degassing of the cryostat when heated around 20 K. 

\section{Quasiparticle dynamics}
\begin{figure}
\centering\includegraphics[width=1\columnwidth]{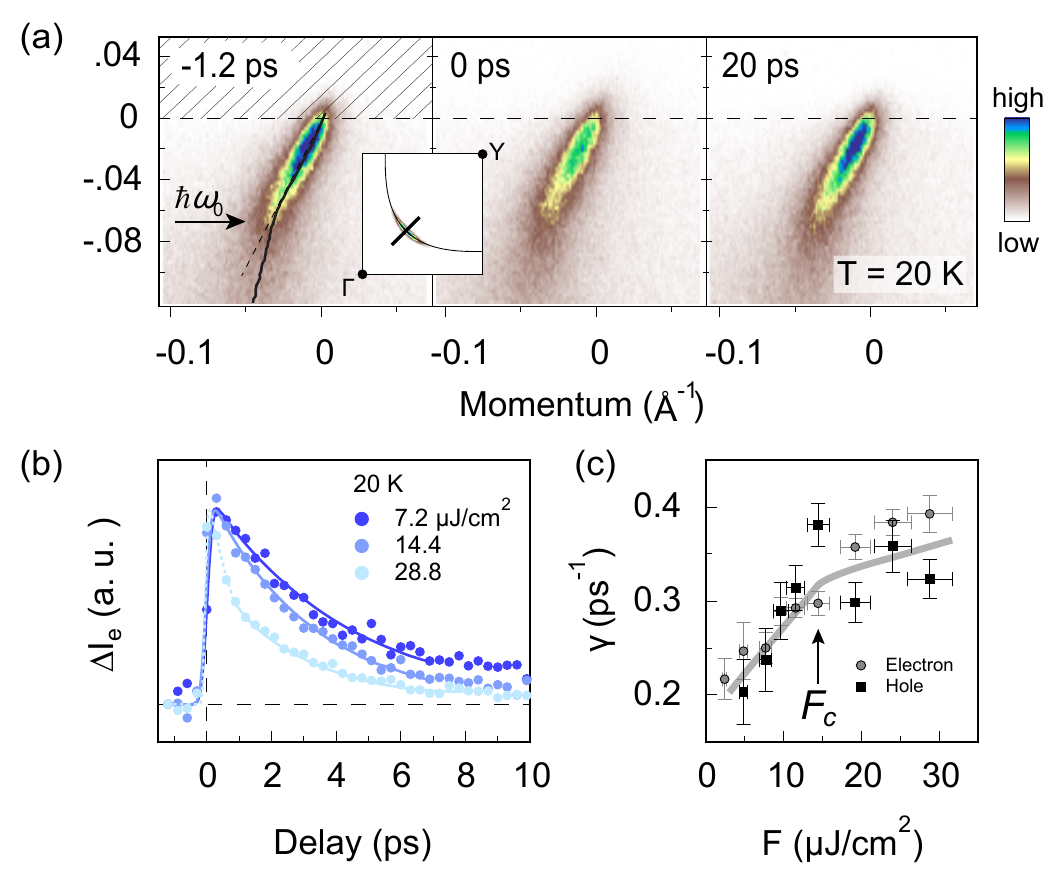}
\caption{
Dynamical evolution of time-resolved ARPES spectra along a nodal cut ($\Gamma$(0, 0) -- Y($\pi$, $\pi$) direction) of underdoped Bi2212 ($T_c=78$ K).
(a) ARPES dispersions: at delay time -1.2, 0, and 20 ps measured at 20 K with pump fluence 14.4 \uJcm. The black solid curve represents the equilibrium electronic dispersion.
(b) Electron-like quasiparticle recombination dynamics for different pump fluences at 20 K. The $\Delta{I}_e$ is obtained by integrating photoemission intensity across the hatched area shown in (a) and then subtracting the integral of the equilibrium intensity. Decay curves are normalized to the same amplitudes.
(c) The non-equilibrium quasiparticle decay rate as a function of pump fluence ($F$) for both electron and hole parts at 20 K.
}
\label{fig1}
\end{figure}

Figure~\ref{fig1} shows results for the UD78K sample.
At a base temperature of 20 K, the equilibrium spectrum ($t=-1.2$ ps) shows a well-known dispersion kink at binding energy $\hbar\omega_0\approx70$ meV as marked by the arrow in Fig.~\ref{fig1}(a)\cite{Lanzara2001}.
As the pump pulse strikes the sample ($t=0$ ps), the intensity of the ARPES spectrum between the Fermi level and the kink energy is suppressed.
It returns to the equilibrium value after 20 ps.
Such evolution of the transient ARPES spectrum is similar to that of optimally doped Bi2212 as reported in Refs.~\onlinecite{Graf2011} and \onlinecite{Smallwood2012}. 
To illustrate the recovery process, Fig.~\ref{fig1}(b) shows the change in integrated ARPES intensity above $E_F$ ($\Delta{I}_\text{e}$, see hatched region in Fig. \ref{fig1}(a)), with the response at different fluences normalized to the same amplitudes.
In the superconducting state, the recovery rate of the non-equilibrium state increases linearly with fluence (Fig.~\ref{fig1}(c)) in a manner similar to that observed in optimally doped Bi2212\cite{Smallwood2012}, suggesting bimolecular recombination\cite{Gedik2004,Kaindl2005}.
As the fluence approaches a critical fluence $F_c$ ($\sim$13 \uJcm), the decay rate undergoes a change in slope,  marking the onset of different recombination processes for quasiparticles.
A similar but higher critical fluence was found in a time-resolved optical reflectivity study\cite{Coslovich2011a}; the difference in thresholds may be because reflectivity measurements probe more of the bulk than ARPES.
Such a fluence threshold is consistent with an observation of the full closure of the superconducting gap at a similar critical fluence in optimally doped Bi2212\cite{Smallwood2013}, and thus we identify it as the likely fluence where all Cooper pairs have been destroyed. The identification is also consistent with a simple back-of-the-envelope calculation of the fluence needed to break apart all Copper pairs.
Indeed, using a superconducting coherence length of $\sim$15 \AA\cite{Mourachkine2002} and a penetration depth of $\sim$100 nm\cite{Hwang2007} (for 1.48 eV photons) in Bi2212, at $F_c$ the laser deposits $\sim$0.25 meV/\AA$^2$ of energy in the top copper oxygen plane, or $\sim$55 meV per coherence area. This is on the order of the energy gap in Bi2212.
We note that a portion of the pulse energy may be transferred to phonons in the initial response, but it should not significantly affect our estimation because the initial non-thermal relaxation is dominated by electron-electron scattering\cite{Basov2011}.
Hence at the critical fluence the laser deposits just enough energy to completely break all Cooper pairs.

As the temperature increases, thermally excited quasiparticles begin to dominate the recombination dynamics\cite{Gedik2004}, making it harder to isolate the contribution of photoexcited non-equilibrium quasiparticles.
This limitation can be overcome by looking at the signal at $t=0$ (averaged over $\sim$300 fs because of time resolution), as the thermally excited quasiparticle have only negligible impact on the initial excited population.
The temperature-dependent non-equilibrium quasiparticle population at $t=0$ is studied in following section.

\section{Fluence and temperature dependence}
\begin{figure*}
\centering\includegraphics[width=1.8\columnwidth]{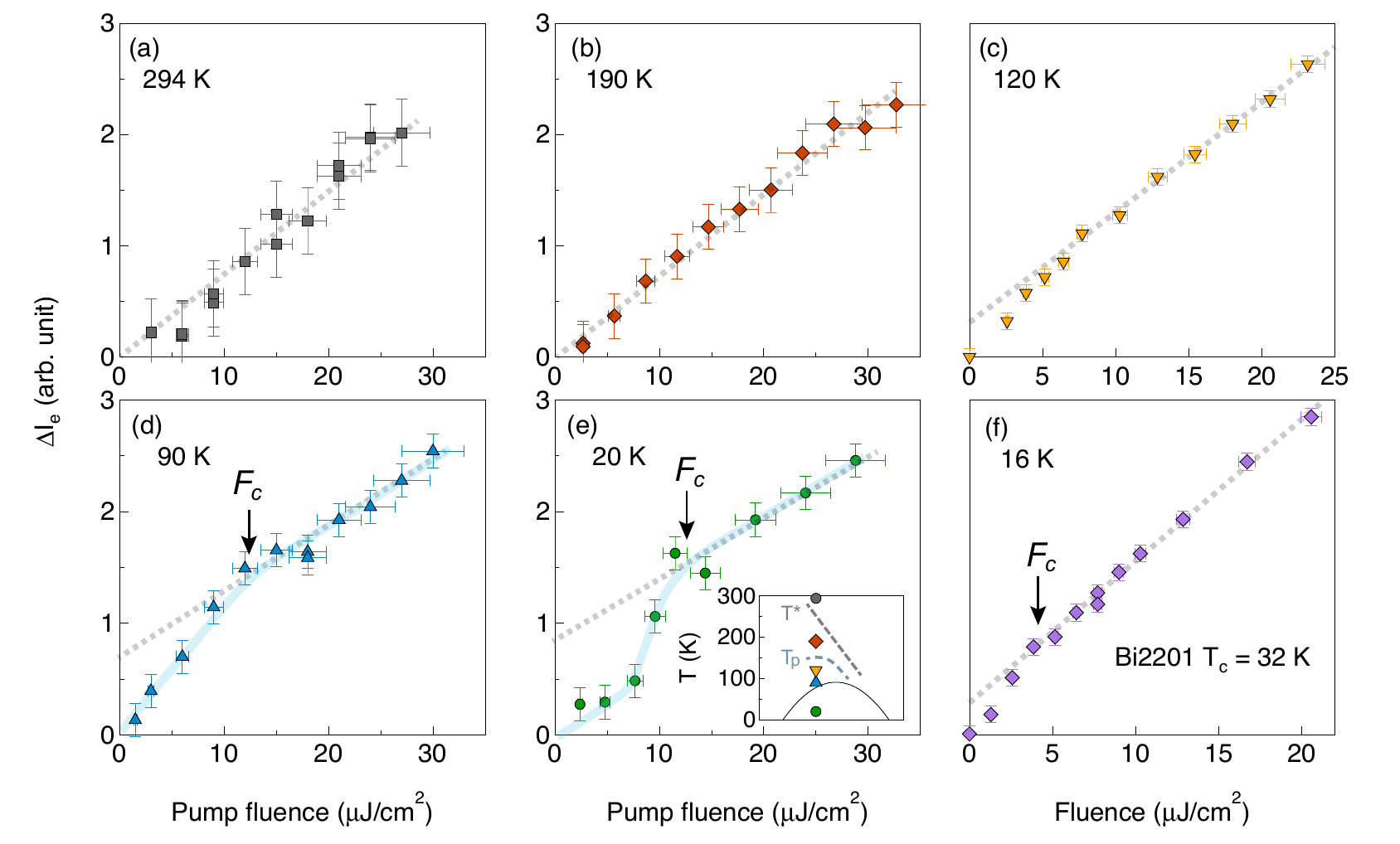}
\caption{
Nodal fluence dependence of the initial quasiparticle excitation density at $t=0$ in underdoped Bi2212 (${T}_c=$ 78 K).
(a) $\Delta{I}_e$ at a delay time of 0 ps measured at 294 K. (b--e) Same as (a) but measured at equilibrium temperature 190 K, 120 K, 90 K, and 20 K.  As indicated in the inset of (d), the measurements correspond to equilibrium temperatures above $T^*$ (294 K), slightly below $T^*$ (190 K), slightly above $T_c$ (90 K), and far below $T_c$ (20 K).
(f) Similar measurement at on an overdoped Bi2201 sample ($T_c=32$ K) at $T=16$ K.
The bold lines are guides to the eye, and the dashed bold lines are linear fits to the high fluence data. The black arrows mark the critical fluence where the slope of the curves changes.
}
\label{fig2}
\end{figure*}

Figure~\ref{fig2} shows $\Delta I_e(t=0)$ as a function of fluence and temperature for an underdoped sample.
Two distinct regimes can be identified from the data:
a high temperature regime, which extends from the normal state (panel (a)) well into the pseudogap state (panel (b), $T^*\approx220$ K\cite{Oda1997,Chatterjee2011}), where $\Delta{I}_e$ is linear in fluence and extrapolates to 0 at $F=0$ (see dashed lines)); and a low temperature regime, which sets in above $T_c$ (panels (b) and (c)) and persists into the superconducting state (panel (d)), where $\Delta I_e$ is clearly not linear in fluence, with high-fluence values of $\Delta I_e$ extrapolating to a positive $y$-intercept at $F=0$.
The departure from linearity at low temperature occurs at the critical fluence ($F_c=13 \pm 3$ \uJcm) identified in Fig.~\ref{fig1}(e).
According to the interpretation that $F_c$ marks the threshold above which no Cooper pairs exist, we observe a slight decrease of $F_c$ from 13 \uJcm (panels (d) and (e)) to 8 \uJcm as the temperature increases to 120 K (panel (c)).
Also, in a single layer Bi2201 sample, we observe a lower $F_c$ by a factor of 3 (panel (f)).
These observations are consistent with the interpretation of $F_c$ and the fact that it should scale with the pairing strength, which gets weaker at higher temperatures, and is weaker in Bi2201 than in Bi2212.
However, the observation of two distinct temperature regimes in the UD78K sample hints at an intermediate temperature scale between $T_c$ and $T^*$, which, given the similar dynamics as in the superconducting state, is likely related to the onset of superconducting fluctuations.
In line with this observation, a recent ARPES experiment has shown that the antinodal spectral function is sensitive to the onset of superconducting pairing fluctuations at an intermediate temperature scale between $T_c$ and $T^*$\cite{Kondo2011}.

\begin{figure}
\centering\includegraphics[width=1\columnwidth]{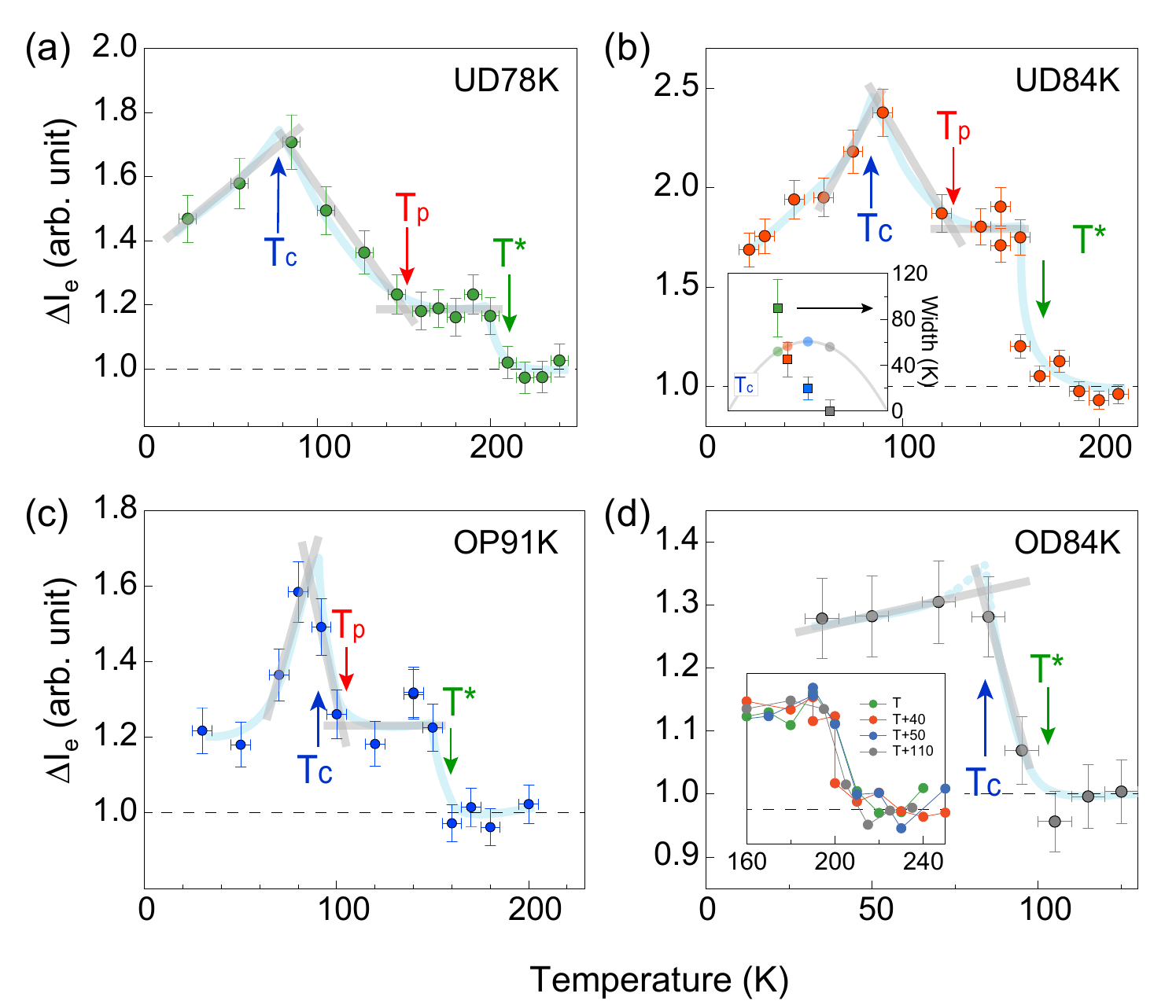}
\caption{
Nodal temperature dependence of the initial quasiparticle excitation density at a fixed fluence of 14.4 ${\mu}$J/cm${^{2}}$ for various dopings of Bi2212.
(a) Underdoped ($T_c=78$  K) sample. $\Delta{I}_e$ at zero delay time, defined as in Fig.~\ref{fig1}, and normalized to the height above $T^*$.
(b--d) Same as (a) for an underdoped $T_c=84$  K sample, an optimally doped $T_c=91$  K sample, and an overdoped $T_c=84$  K sample.
The bold cyan lines are guides to the eyes.
$\Delta{I_e}$ at different temperature are normalized to the same probe fluence.
Inset in (b) shows the width of the peak-like feature around $T_c$ as a function of doping\cite{SM}.
Inset in (d) is the $\Delta{I}_e$ vs temperature for the four dopings of sample, which are scaled to the same amplitudes at $T^*$.
Error bars are taken to be the maximum $|\Delta I(t))|$ for $t < 0$.
}
\label{fig3}
\end{figure}

In Fig.~\ref{fig3} we show detailed temperature and doping dependence of $\Delta{I}_e$ at $t=0$ for a fixed pump fluence (14.4 \uJcm, slightly above the critical fluence).
We note that the reflectance of Bi2212 has little temperature dependence from 4 K to 300 K at photon energy 1.48 eV (variation is less than 1$\%$)\cite{Tanaka1999}, guaranteeing that the pump fluence applied on the sample is unchanged during the temperature-dependent measurements.
A common feature to all the dopings is a sharp step in $\Delta{I}_e$ at $T^*$, which coincides with the pseudogap temperature determined from transport experiments\cite{Oda1997} and from momentum-resolved experiments looking at antinodal quasiparticles\cite{Chatterjee2011,Kondo2011}.
The sharp step is still observed above $T_c$ for an overdoped sample (Fig.~\ref{fig3}(d)), suggesting that the pseudogap temperature $T^*$ can be defined even for this overdoped sample, consistent with recent reports by equilibrium ARPES on antinodal quasiparticles\cite{Chatterjee2011,Vishik2012}.
The step indicates that non-equilibrium electronic states below $T^*$ result in a larger contribution to $\Delta{I}_e$ at the node, even though there is no gap at the node itself as generally believed.

For the underdoped and optimally doped samples (panels (a)--(c)), further cooling reveals a distinctive peak-like feature in $\Delta{I}_e$ centered at $T_c$, and a plateau in $\Delta{I}_e$ above $T_c$ that is bounded on the high temperature side by $T^*$ and on the low temperature side by a temperature $T_p$, defined as the temperature below which $\Delta{I}_e$ begins to rise.
The peak-like feature at $T_c$ is reminiscent of a variety of observables that diverge in the vicinity of $T_c$ as a result of phase fluctuations. These include, but are not limited to, $\lambda$-shaped anomalies in the temperature dependence of thermal expansivity coefficients in YBCO\cite{ Meingast2001}, and a theoretical divergence of relaxation time at the critical temperature in random-field Ising systems\cite{Fisher1986}.
Thus it is reasonable to infer that the peak in $\Delta{I}_e$ around $T_c$ is a measure of phase fluctuations.
Further confirmation comes from the sharpening of the peak as the doping increases, reflecting a narrowing of the Ginzburg window\cite{Ginzburg1960} (see the width of this peak-like feature as a function of doping in the inset of Fig.~\ref{fig3}(b)) .
If the peak at $T_c$ is associated with the presence of superconducting phase fluctuations, then $T_p$, the end of the peak feature, should be identified with the onset of such fluctuations.
This scenario is further supported by the good agreement with the Nernst effect temperature\cite{Wang2006}, identified as the onset of superconducting fluctuations in cuprate superconductors and the $T_p$ measured in this experiment is along the nodal direction.
The absence of $T_p$ in the overdoped sample (panel (d)) implies that the superconducting transition is determined by Cooper pair formation rather than phase fluctuations.
The non-equilibrium spectra in Figs.~\ref{fig3}(a) and (b) do not saturate at the lowest temperature in our measurements for the two underdoped sample, indicating that fluctuating uncondensed pairs exist farther below $T_c$ than on the overdoped side.
Our result also demonstrates that below $T_p$, the breaking of uncondensed Cooper pairs by pumping dominates $\Delta{I}_e$ at $t=0$.

The features at $T_c$, $T_p$, and $T^*$ follow distinct trends as a function of doping.
For all dopings, fitting the rise of the step at $T^*$ using an error function yields a temperature smearing less than 10 K (inset of panel (d)).
The abrupt onset at $T^*$ is reminiscent of time-resolved reflectivity measurements on Bi2201, where a similar behavior was associated with $T^*$, marking the onset of a phase transition\cite{He2011}.
The fact that $\Delta{I}_e$ at $t=0$ shows the same critical fluence in both the superconducting state and between $T_c$ and $T_p$ (Figs.~\ref{fig2}(c)--(d)), demonstrates that the response of the electronic state to the pump pulse between $T_p$ and $T_c$ is similar to that in the superconducting state.
In contrast, the absence of a critical fluence above $T_p$ in Fig.~\ref{fig2} shows the different response between states below $T_p$ and states below $T^*$, indicating the different underlying interactions of the pseudogap and the superconducting state.

\begin{figure}[htbp]\centering\includegraphics[width=0.8\columnwidth]{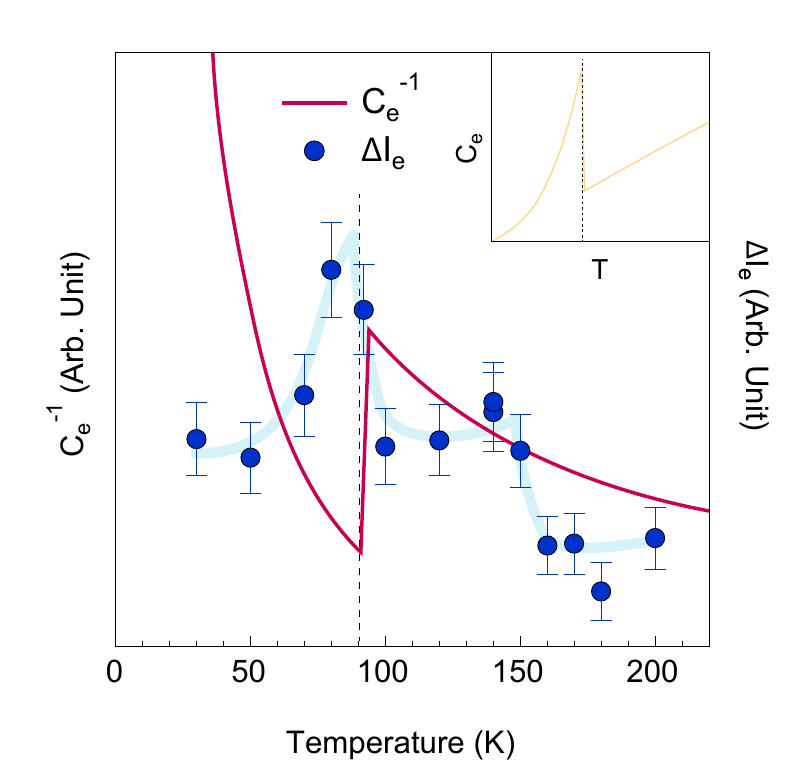}
  \caption{
Comparison between initial nonequilibrium electrons $\Delta{I}_\text{e}$ and simulated electron heat capacity $C_\text{e}$ as a function of temperature. The inset shows the $C_\text{e}$ as a function of temperature in a superconductor. 
  }
  \label{fig4}
\end{figure}

We note that the step at $T^*$ and peak-like feature at $T_c$ cannot be attributed to thermal effects.
Indeed, in a thermal model the initial non-equilibrium electron population $\Delta{I_\text{e}}^{simu}$ is proportional to the energy integral of a Fermi-Dirac distribution between the chemical potential and infinity, which is in turn proportional to the electronic temperature $ T_\text{e}$:
\begin{eqnarray}
\label{equ:Gain1}
{I}_\text{e}^{simu}&\propto&\int_0^{+\infty}\frac{1}{e^{\omega/k_BT_\text{e}}+1}d\omega\\
\label{equ:Gain2}
&\propto&
k_BT_\text{e}\int_0^{+\infty}\frac{1}{e^x+1}dx\\
\label{equ:Gain3}
&\propto&{T_\text{e}}
\end{eqnarray}
It immediately follows that the change in the electronic spectral weight ($\Delta{I_\text{e}}$) is proportional to the change of electronic temperature $\Delta{T_\text{e}}$ and independent of the equilibrium temperature.

If one assumes that the energy of the pump pulse $\Delta{Q}$ is mainly absorbed to heat the electrons\cite{Basov2011}, then
\begin{equation}
d{Q}=d{T_\text{e}}\cdot{C_\text{e}}
\label{differential}
\end{equation}
where $C_\text{e}$ is the electronic heat capacity. According then to Eqs.~\ref{equ:Gain1} to \ref{equ:Gain3}, we expect that $\Delta{I_\text{e}}$ is inversely proportional to the specific heat capacity $C_\text{e}$.  In Fig. \ref{fig4} we use a simple model to simulate the electron heat capacity $C_\text{e}$ with $C_\text{e}=AT^3$ below $T_\text{c}$ and $C_\text{e}=BT$ above $T_\text{c}$ ($A$ and $B$ are constants).
The direct comparison between $C_\text{e}$ and $\Delta{I_\text{e}}$ shown in Fig. \ref{fig4} clearly shows that the initial spectral gain above the Fermi level cannot be attributed to a simple thermal effect.  Indeed, while the experimental value of $\Delta{I_\text{e}}(t)$ shows a peak-like feature around $T_\text{c}$, the model predicts a dip in $C_\text{e}^{-1}$ at the same temperature.
The basic shape of this dip feature at $T_\text{c}$ is robust, even after integrating Eq.~\ref{differential} to account for realistically finite values of $\Delta{I_\text{e}}$ and $\Delta{T_\text{e}}$.

\section{Phase diagram and discussion}
\begin{figure}
\centering\includegraphics[width=1\columnwidth]{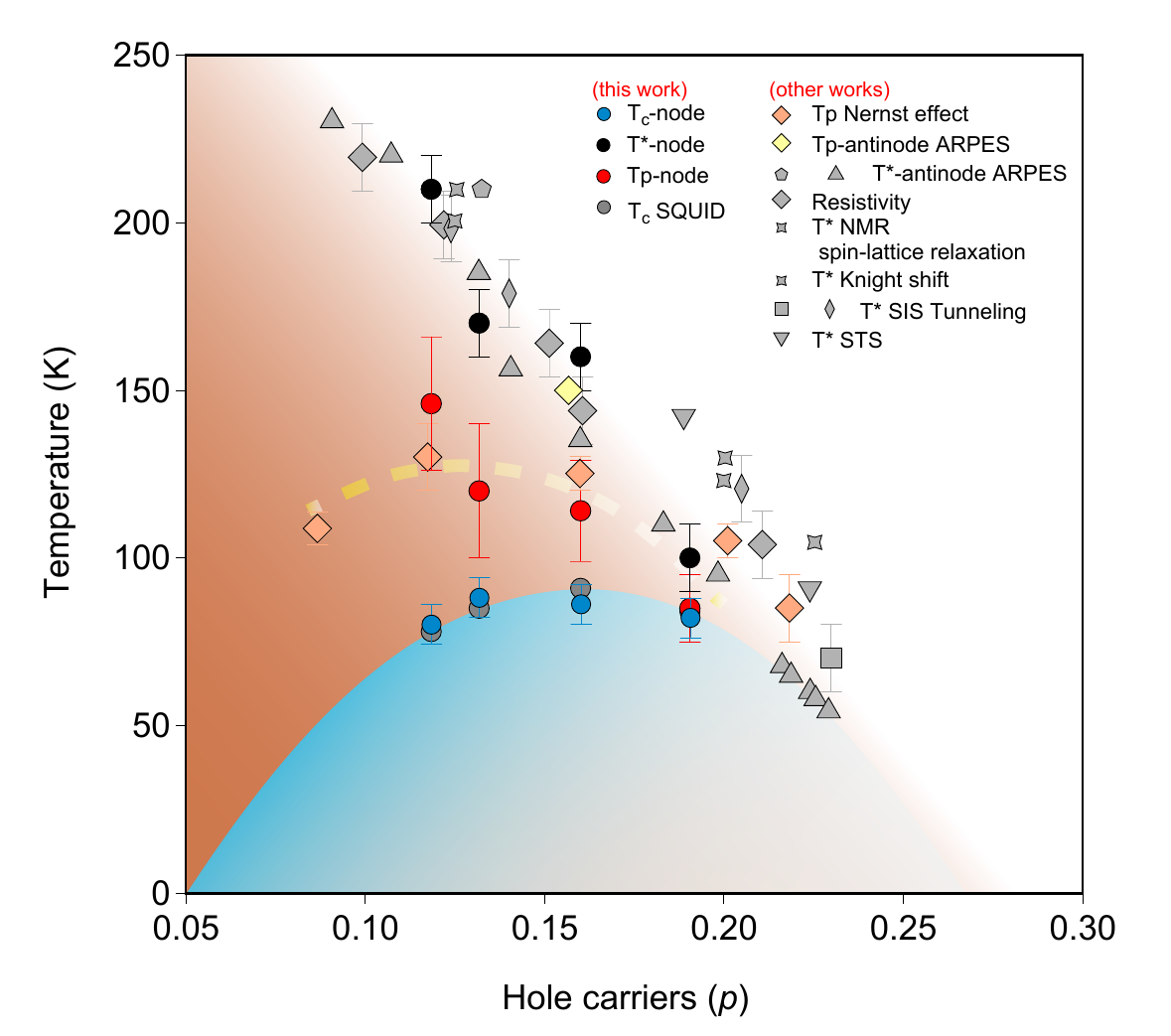}
\caption{
Nodal phase diagram of Bi2212.
The superconducting critical temperature $T_c$-node, superconducting fluctuation onset temperature $T_p$-node and pseudogap temperature $T^*$-node of the four different dopings are plotted. $T_c$-node (blue filled circles), $T^*$-node (black filled circles) and $T_p$-node (red filled circles) are determined by the transition and onset temperature in Fig.~\ref{fig3}; $T_c$ (gray filled circles) is derived from SQUID measurements.
$T_p$ from Nernst signal measurements (pink diamonds)\cite{Wang2006} and off-nodal ARPES measurement (yellow diamond)\cite{Kondo2011}, as well as $T^*$ (gray data points) from ARPES\cite{Chatterjee2011,Kondo2011}, resistivity\cite{Oda1997}, nuclear spin-lattice relaxation\cite{Ishida1998}, the Knight shift\cite{Ishida1998}, SIS tunneling\cite{Ozyuzer2002,Dipasupil2002} and STS\cite{Gomes2007} are also plotted. The hole carrier concentration $p$ of each sample is calculated by the Presland-Tallon equation $T_c/T_c^{max}=1-82.6(p-0.16)^2$\cite{Presland1991}.
}
\label{fig5}
\end{figure}

In Fig.~\ref{fig5}, we summarize the temperatures identified for the nodal transient spectra $T_c$, $T^*$, and $T_p$, in the form of a phase diagram, and compare them with similar temperature scales reported in the literature.
The most important implications of this comparison are
(1) the position of the peak-like feature ($T_c$-node) in Fig.~\ref{fig3} matches the superconducting critical temperature $T_c$ measured by SQUID;
(2) $T_p$-node coincides with the onset temperature of superconducting phase fluctuations measured by Nernst effect\cite{Wang2006}; and (3) $T^*$-node coincides with the pseudogap temperature extracted from various momentum integrated probes\cite{Oda1997,Ishida1998,Ozyuzer2002,Dipasupil2002,Gomes2007} and from ARPES along the antinodal direction ($T^*$-antinode)\cite{Chatterjee2011,Kondo2011,Vishik2012}.

Revealing the presence of these energy scales in the spectral function of ungapped quasiparticles disrupts the conventional view that $T_c$, $T_p$, and $T^*$ are only associated with gapped antinodal states.
The signature at $T^*$ is perhaps the most surprising of these, as the most popular explanation for the pseudogap phase at present is that it is associated with onset of charge ordering with a nesting vector along the ($\pi$,0) direction, resulting therefore in strongly suppressed antinodal electronic states\cite{Fu2006}.
The signatures of $T^*$ and $T_c$ in the ($\pi$,$\pi$) direction may indicate that the gapped antinodal non-equilibrium quasiparticles in both of the pseudogap and superconducting states can be scattered to nodal region via exchanging a momentum with other excitations in a very short time scale.
Theoretical and experimental studies seems to argue against this possibility as this type of scattering is predicted to be pair-breaking\cite{Howell2004,Gedik2005,Cortes2011}, leaving a far more exciting possibility also exists, namely, that nodal electronic states intrinsically play the same role in shaping the pseudogap and superconducting states on the phase diagram as generally believed in antinodal states.

In summary, we have revealed a strong response of non-equilibrium nodal ungapped quasiparticles to both the superconducting and pseudogap states. 
A new phase diagram for nodal quasiparticles results from these data,  similar to the one widely discussed for gapped antinodal quasiparticles, where the pseudogap temperature $T^*$ gradually merges with the superconducting transition temperature on the overdoped side of the phase diagram, and an intermediate temperature scale $T_p$ associated with uncondensed Cooper pairs sets in between $T_c$ and $T^*$.
These results highlight the important role that nodal quasiparticles play for cuprate superconductivity, as well as the different electronic natures of the pseudogap and superconducting transition.

\begin{acknowledgments}
We thank J. Orenstein, S. Kivelson and P. Phillips for useful discussions, and J. S. Wen, J. Zhao and R. J. Birgeneau for SQUID measurements.
This work was supported by Berkeley Lab's program on Quantum Materials, funded by the U.\ S.\ Department of Energy, Office of Science, Office of Basic Energy Sciences, Materials Sciences and Engineering Division, under Contract No.\ DE-AC02-05CH11231.
\end{acknowledgments}

%

\end{document}